\documentclass[10pt]{article}
\usepackage{natbib}
\usepackage{amssymb,amsmath}
\usepackage{graphicx}
\usepackage{color}
\usepackage{hyperref}
\hypersetup{
	colorlinks=true,
	linkcolor=black,
	filecolor=red,      
	citecolor=blue,
}
\usepackage[dvipsnames]{xcolor}
\definecolor{myblue}{RGB}{31,119,180}
\definecolor{myorange}{RGB}{255,127,14}
\definecolor{mymagenta}{RGB}{255,0,255}
\definecolor{mygreen}{RGB}{0,128,0}
\usepackage{curves,multicol}
\usepackage{geometry}
\usepackage{times}
\usepackage{bm}
\usepackage{ulem}
\usepackage{ucs}
\usepackage[utf8x]{inputenc}
\usepackage[intoc]{nomencl}
\usepackage{float}
\usepackage{mathrsfs}
\usepackage{cleveref}
\usepackage{units}
\usepackage{titlesec}
\usepackage[labelfont=bf]{caption}
\usepackage{lipsum}   
\usepackage[none]{hyphenat} 
\usepackage{makecell}
\usepackage{tikz}

\usepackage{scalerel,stackengine}
\stackMath
\newcommand\reallywidehat[1]{%
	\savestack{\tmpbox}{\stretchto{%
			\scaleto{%
				\scalerel*[\widthof{\ensuremath{#1}}]{\kern-.6pt\bigwedge\kern-.6pt}%
				{\rule[-\textheight/2]{1ex}{\textheight}}%WIDTH-LIMITED BIG WEDGE
			}{\textheight}% 
		}{0.5ex}}%
	\stackon[1pt]{#1}{\tmpbox}%
}
\crefname{equation}{Equation}{Equations}
\Crefname{equation}{Equation}{Equations}
\Crefname{figure}{Figure}{Figures}
\crefname{table}{Table}{Tables}
\Crefname{tabular}{Table}{Tables}

\def\dashdotted{\xleaders\hbox to 1em{$- \cdot$}\hfill $-$}

\makeatletter

\renewcommand\@biblabel[1]{}

\renewcommand{\section}{\@startsection
{section}%                   % the name
{1}%                         % the level
{0mm}%                       % the indent
{-\baselineskip}%            % the before skip
{0.5\baselineskip}%          % the after skip
{\normalfont\bfseries\MakeUppercase}} % the style

\renewcommand{\subsection}{\@startsection
{subsection}%                   % the name
{2}%                         % the level
{0mm}%                       % the indent
{0.5\baselineskip}%            % the before skip
{0.25\baselineskip}%          % the after skip
{\bfseries\normalsize}} % the style

\makeatother
\begin{document}
\sloppy
\pagenumbering{arabic}
\setcounter{secnumdepth}{-1} 

\vspace*{-1.0cm}
\begin{flushright} \vbox{
34th Symposium on Naval Hydrodynamics\\
Washington, DC, USA, 26 June -- 1 July 2022}
\end{flushright}

\vskip0.65cm
\begin{center}
\textbf{\LARGE
Evolution of Wave Characteristics during \\Wind-Wave Generation\\[0.35cm]
}

\Large Tianyi Li and Lian Shen\\ 

(St. Anthony Falls Laboratory and Department of Mechanical Engineering, University of Minnesota, Minneapolis, MN 55455, USA)\\
\vspace*{0.25cm}

\end{center}

\begin{multicols*}{2}

\section{Abstract}

The generation of ocean surface waves by wind is a classic fluid mechanics problem whose theoretical study dates back to 1957, when the two seminal papers by Phillips and Miles were published in the incipient Journal of Fluid Mechanics.  Comprehensively understanding the mechanism of wind-wave generation is of crucial importance to many naval applications.  Here, we present the recent developments of our work on this profound and long-standing problem.  We perform a wave phase-resolved direct numerical simulation (DNS) of the turbulent airflow over an initially calm water surface that responds dynamically to the wind forcing.  Our simulation reveals the evolution of the surface wave statistics over the entire wind-wave generation process.  As the wave fluctuations grow with time under the turbulent wind forcing, the temporal growth behavior of the surface elevation variance transitions from linear growth to exponential growth.  Moreover, surface waves exhibit nonlinearity at the late stage of simulation when the wave amplitude is significant enough to alter the airflow field.  Based on this rich DNS dataset, we systematically study the underlying mechanism and report several advances in the theoretical analysis of this problem.

\section{Introduction}\label{sec:intro}
The ocean is never still, and surface waves propagate day and night. How ocean waves are generated has remained a fundamental question for decades. 
In a collection of surveys published in 1956 to commemorate G. I. Taylor's 70th birthday, \cite{ursell1956wave} highlighted the need to understand the mechanisms responsible for generating wind-waves.  
Soon thereafter in 1957, \cite{phillips1957generation} and \cite{miles1957generation} established two theories on the generation mechanisms of wind-waves; both seminal papers were published in the Journal of Fluid Mechanics almost immediately after its founding in 1956.  
\cite{phillips1957generation} postulated that turbulent airflow pressure fluctuations at the ocean surface are the main source of surface waves; 
moreover, he proposed a stochastic model on the basis of an asymptotic analysis showing that the surface wave variance grows linearly with time in the principal stage. 
In contrast, \cite{miles1957generation} studied the wind-wave generation problem from the perspective of stability analysis and developed a quasilinear model demonstrating that the wave amplitude grows exponentially with time.
It is commonly believed that the Phillips theory describes the early stage of wave generation, whereas the Miles theory describes wave dynamics in the late stage of wave generation.  We refer the readers to \cite{li2022principal} for a review of the wind-wave generation literature.
At present, systematic research on the wind-wave generation process is still required to validate or modify these classic wind-wave generation theories. 
However, owing to the complexity of the turbulence motions occurring near the air--water interface, the need for high-resolution simulations and experiments continues to impose considerable demands on the available computing power and experimental techniques.

The Phillips theory \citep{phillips1957generation} assumes that the air--water interface is distorted by quasisteady turbulent air pressure fluctuations.
The space--time characteristics of these turbulent pressure fluctuations delineate a principal stage of wind-wave generation when the variance of the surface elevation $\eta$ is proportional to the time $t$:
\begin{align}
	\langle\eta^2\rangle\sim \frac{\langle p^2\rangle t}{2\sqrt{2}\rho^{w2}gU_c},\label{eq:phillips}
\end{align}
where $p$ denotes the air pressure at the air--water interface, $\rho^w$ denotes the water density, $U_c$ denotes the convection velocity of pressure fluctuations, $\langle\,\cdot\,\rangle$ denotes a spatial averaging operation, and `$\sim$' signifies proportionality.
The existence of this principal wind-wave generation stage was first confirmed numerically by \citet{lin2008direct} and experimentally by \citet{zavadsky2017water}.
However, \cref{eq:phillips} could not provide a quantitative expression of the surface elevation variance because the convection velocity $U_c$ alone is not sufficient to model the space--time characteristics of the convection of turbulent pressure fluctuations.  
Recently, \cite{li2022principal} introduced another turbulence quantity, namely, the sweeping velocity of turbulent pressure fluctuations \citep[see][]{he2006elliptic,wilczek2012wave}, to the framework of the Phillips theory and obtained a quantitative prediction of the surface elevation variance over time.

As the surface elevation grows, the motions of surface waves exert a gradually increasing influence on the airflow; this results in the generation of wave-coherent airflow structures.
Accordingly, \cite{miles1957generation} proposed that when the airflow velocity at a certain height equals the surface wave velocity, a critical-layer instability arises, and the surface elevation grows exponentially with time. 
The dimensionless wave growth rate parameter introduced by \citet{miles1957generation} is defined as
\begin{equation}
	\beta=\frac{1}{\omega E}\frac{\mathrm{d}E}{\mathrm{d}t},\label{eq:beta}
\end{equation}
where $E$ denotes the wave energy and $\omega$ denotes the wave frequency. The constant value of the wave growth rate $\beta$ indicates that the total wavefield energy $E$ grows exponentially with time $t$.  
The wave growth rate $\beta$ is a function of the wave age $c/u_\tau$, where $c$ and $u_\tau$ denote the wave phase velocity and airflow friction velocity, respectively.  Variations in the wave growth rate with the wave age have been extensively studied by numerical simulations \citep[e.g.,][]{li2000numerical,sullivan2000simulation,yang2010direct,hao2019wind} and experimental observations \citep[e.g.,][]{mastenbroek1996experimental,grare2013growth}.
When the wave steepness exceeds a certain threshold, wave breaking occurs and enhanced dissipation suppresses the wave growth.
Recently, \citet{geva2022excitation} proposed a fetch-limited growth theory based on the Orr--Sommerfeld equation to explain multi-stage wave growth behaviors till the final quasi-steady state, which agrees well with wave tank experiments.
In particular, the numerical simulations of \citet{lin2008direct} and \citet{li2022principal} revealed a transition stage between the linear growth stage \citep{phillips1957generation} and the exponential growth stage \citep{miles1957generation}. 
In addition, \cite{perrard2019turbulent} proposed a surface elevation transition threshold to trigger the exponential growth of surface waves from a scaling analysis on the experimental data of \citet{paquier2015surface,paquier2016viscosity}, and the concept of this threshold was later supported by the numerical simulation of \citet{li2022principal}.
Nevertheless, additional theoretical advances are necessary to comprehensively understand the transition stage between the wind-wave growth stages described by the Phillips theory and the Miles theory.

In the present study, we analyze the evolution of surface elevation characteristics based on the direct numerical simulation (DNS) database of the entire wind-wave generation process from \citet{li2022principal}.  
\textcolor{black}{In \citet{li2022principal}, the linear growth behavior of the surface elevation variance in the principal stage of the wind-wave generation process was systematically evaluated, and a random sweeping turbulence pressure–wave interaction model was proposed to quantitatively obtain the wave growth rate in the principal stage.  Here, we extend the analysis on the second-order statistics of the surface elevation in \citet{li2022principal} to the arbitrary high-order statistics and related correlation functions, with a focus on their evolutions during the entire wind-wave generation process.}
The high-order statistics of the surface elevation, the anisotropy factor of the surface elevation, and the correlation function between surface elevation gradients and pressure fluctuations are studied in a combined numerical and theoretical approach.

\section{Simulation Setup}\label{sec:simulation_setup}
We numerically study the evolution of wave characteristics during the wind-wave generation process in a horizontally periodic rectangular domain.
Assuming that no wave breaking occurs during wind-wave generation, the entire domain can be divided into a simply connected air domain and a simply connected water domain, which are denoted $\Omega^a$ and $\Omega^w$, respectively.
In both the air domain and the water domain, the following continuity and Navier--Stokes equations govern the fluid motions:
\begin{subequations}\label{eq:NSE}
	\begin{align}
		&\frac{\partial u^a_i}{\partial x_i^a}=0, & \mathrm{in}\,\,\Omega^a,\label{eq:NSE1:sub1}\\
		&\frac{\partial u^a_i}{\partial t}+\frac{\partial (u^a_i u^a_j)}{\partial x_j^a}=-\frac{1}{\rho^a}\frac{\partial p^a}{\partial x_i^a}+\nu^a\frac{\partial^2 u^a_i}{\partial x_j^{a2}}, & \mathrm{in}\,\,\Omega^a,\label{eq:NSE1:sub2}\\
		&\frac{\partial u^w_i}{\partial x_i^w}=0, & \mathrm{in}\,\,\Omega^w,\label{eq:NSE1:sub3}\\
		&\frac{\partial u^w_i}{\partial t}+\frac{\partial (u^w_i u^w_j)}{\partial x_j^w}=-\frac{1}{\rho^w}\frac{\partial p^w}{\partial x_i^w}+\nu^w\frac{\partial^2 u^w_i}{\partial x_j^{w2}}, & \mathrm{in}\,\,\Omega^w.\label{eq:NSE1:sub4}
	\end{align}
\end{subequations}
In these equations, the coordinates $x_i$ with $i=1,2,3$ or $(x,y,z)$ denote the streamwise, spanwise and vertical directions, respectively, while $u_i$ with $i=1,2,3$ denotes the velocity components in the corresponding directions.
The superscripts `$a$' and `$w$' denote the variables in the air domain and water domain, respectively.
$\rho$ denotes the density, $p$ denotes the pressure, and $\nu$ denotes the kinematic viscosity.
Both the air and the water domains have dimensions of $L_x\times L_y$, and
the heights of the air and water domains are $H^a$ and $H^w$, respectively, when the air--water interface is flat.
In the present study, we set $(L_x,L_y)=(2\pi,\pi)$ and $H^a=H^w=H=1$.
A streamwise-constant shear stress is applied to the top of the air domain to drive the flow field.
A no-slip boundary condition is imposed on the bottom of the water domain, and
periodic boundary conditions are applied in the horizontal directions.
Before the simulation begins, a fully developed turbulent Couette airflow is imposed in the air domain. 
At $t=0$, the water is still, and the air--water interface is flat, but
the subsequent motions of air turbulence during the numerical simulation distort the air--water interface and generate wind-waves.

\begin{figure}[H]
	\centerline{\includegraphics[width=1\columnwidth]{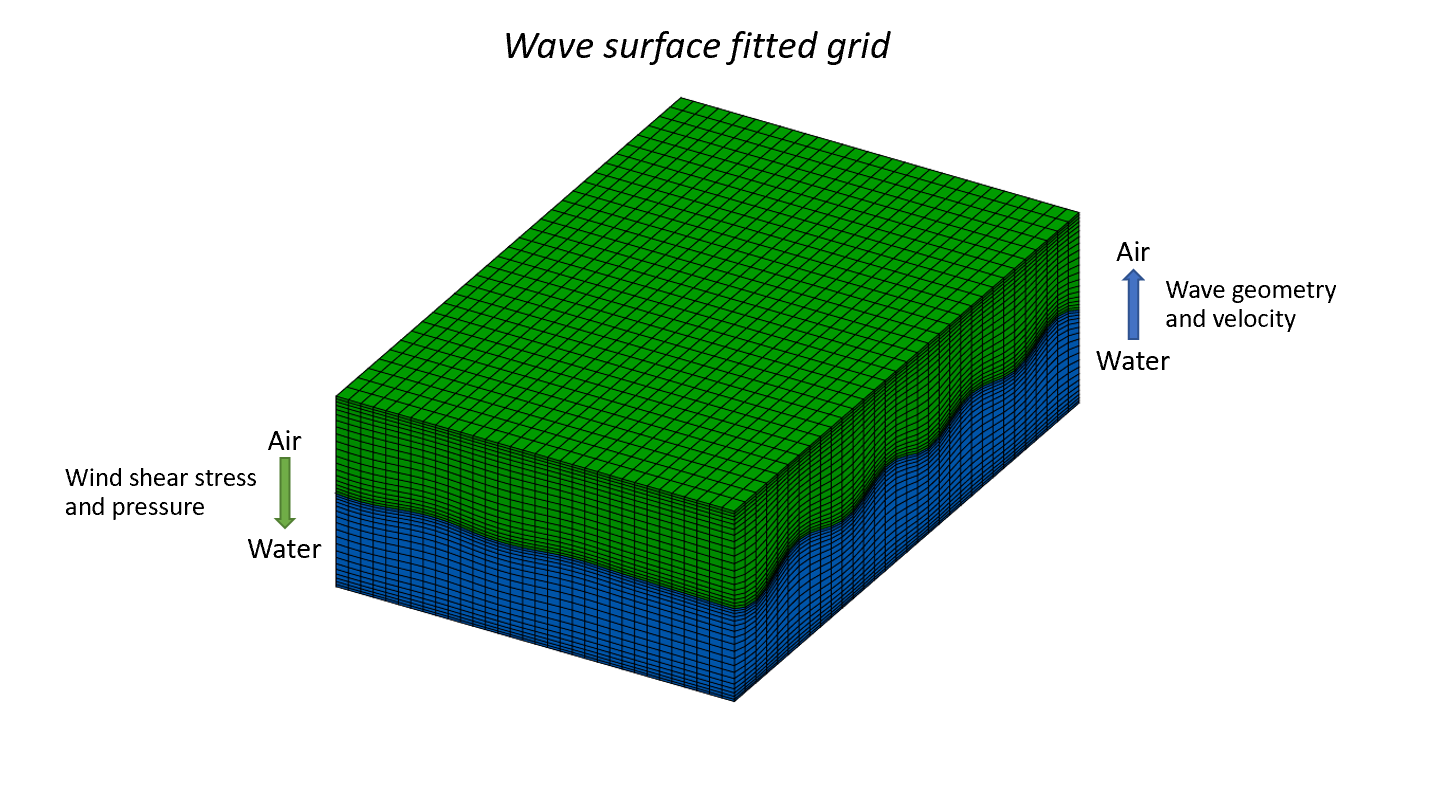}}% Images in 100% size
	\caption{Configuration of the boundary-fitted two-phase flow DNS for the wind-wave generation problem.}
	\label{fig:grid}
\end{figure}
\begin{figure}[H]
	\centerline{\includegraphics[trim={0.0inch 0inch 0  0},clip,width=1\columnwidth]{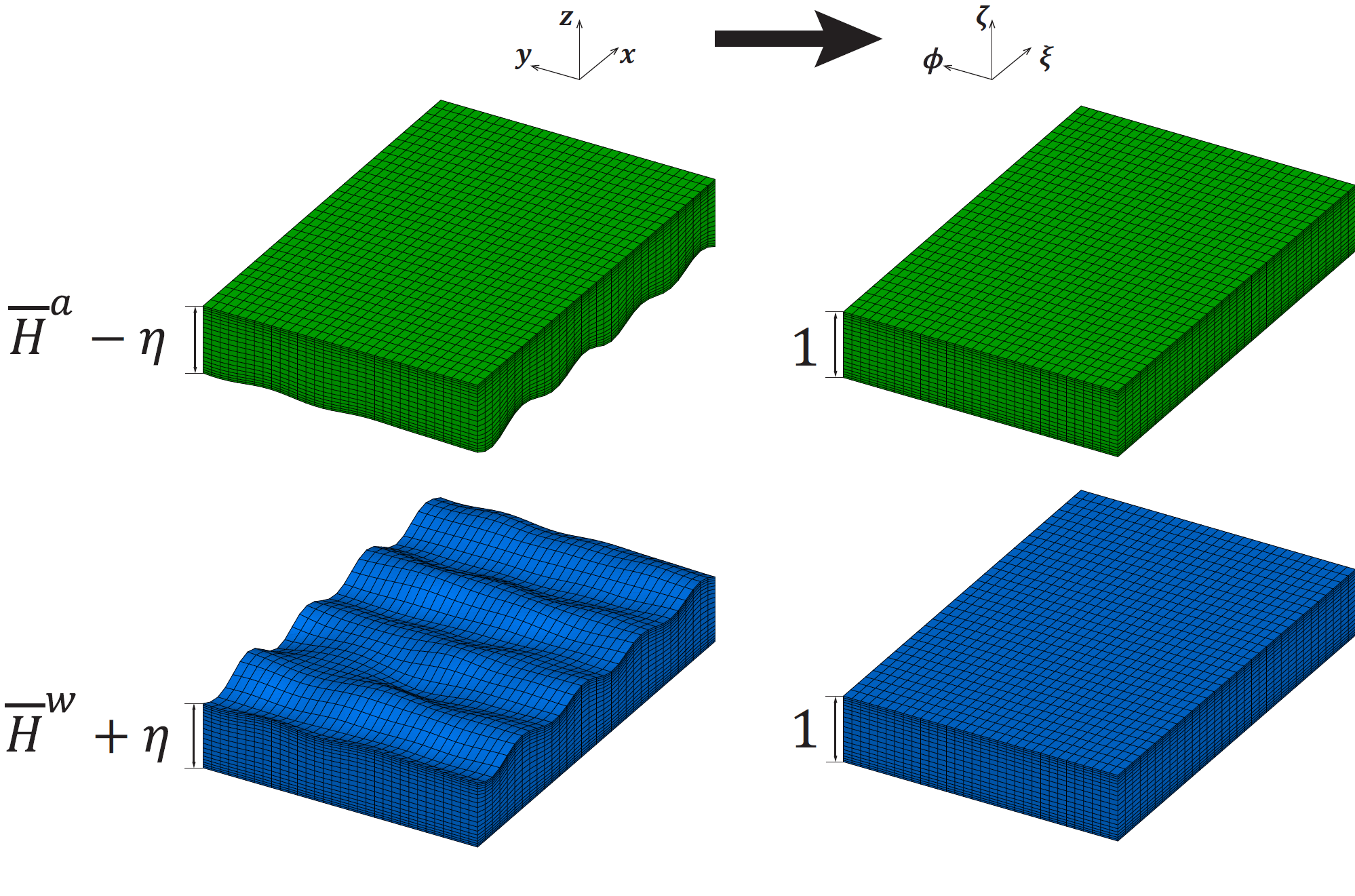}}% Images in 100% size
	\caption{Sketch of coordinate transformation that maps the irregular air and water domain $(x,y,z,t)$ to the corresponding rectangular computational domain $(\xi,\varphi,\zeta,\tau)$.}
	\label{fig:grid2}
\end{figure}
The air domain and water domain are discretized on a dynamically evolving water surface-fitted grid (see \cref{fig:grid}), and the evolution of the air--water interface is explicitly resolved through the nonlinear kinematic and dynamic boundary conditions.
As illustrated in \cref{fig:grid2}, an algebraic mapping is utilized to transform each physical domain $(x,y,z,t)$ into a rectangular computational domain $(\xi,\varphi,\zeta,\tau)$ to effectively capture the exact boundary conditions on the irregular time-varying air--water interface:
\begin{align}
	&\xi^a=x^a,\,\, \varphi^a=y^a,\,\, \zeta^a=\frac{z-\eta(x,y,t)}{-\eta(x,y,t)+{H^a}},\tau^a =t^a,\\
	&\xi^w=x^w,\,\, \varphi^w=y^w,\,\, \zeta^w=\frac{z+H^w}{\eta(x,y,t)+H^w},\tau^w=t^w.
\end{align} 

Next, we transform \cref{eq:NSE1:sub1,eq:NSE1:sub2,eq:NSE1:sub3,eq:NSE1:sub4} from the physical coordinate system $(x,y,z,t)$ to the computational coordinate system $(\xi,\varphi,\zeta,\tau)$ in both the air domain and the water domain, and we obtain the following governing equations in the computational domain:
\begin{subequations}\label{eq:sec03:computational}
	\begin{align}
		&\frac{\partial u^\delta}{\partial \xi^\delta}+\zeta^\delta_x\frac{\partial u^\delta}{\partial\zeta^\delta}+\frac{\partial v^\delta}{\partial \varphi^\delta}+\zeta^\delta_y\frac{\partial v^\delta}{\partial \zeta^\delta}+\zeta^\delta_z\frac{\partial w^\delta}{\partial \zeta^\delta}=0,\label{eq:sec03:computational1}\\
		&\frac{\partial u^\delta}{\partial\tau^\delta}+\zeta^\delta_t\frac{\partial u^\delta}{\partial\zeta^\delta}+\frac{\partial (u^\delta u^\delta)}{\partial\xi^\delta}+\zeta^\delta_x\frac{\partial(u^\delta u^\delta)}{\partial\zeta^\delta}
		+\frac{\partial(u^\delta v^\delta)}{\partial \varphi^\delta}\notag\\
		&+\zeta^\delta_y\frac{\partial(u^\delta v^\delta)}{\partial\zeta^\delta}
		+\zeta^\delta_z\frac{\partial(u^\delta w^\delta)}{\partial\zeta^\delta}=-\frac{1}{\rho^a}\left(\frac{\partial p^\delta}{\partial \xi^\delta}+\zeta^\delta_x\frac{\partial p^\delta}{\partial \zeta^\delta}\right)\notag\\
		&+\nu^\delta\left(
		\frac{\partial^2 u^\delta}{\partial\xi^{\delta2}}+\frac{\partial^2 u^\delta}{\partial\varphi^{\delta2}}+\left(\zeta_x^{\delta2}+\zeta_y^{\delta2}+\zeta_z^{\delta2}\right)\frac{\partial^2 u^\delta}{\partial\zeta^{\delta2}}\right.\notag\\
		&\left.+2\zeta_x^\delta\frac{\partial^2 u^\delta}{\partial\xi^\delta\partial\zeta^\delta}+2\zeta_y^\delta\frac{\partial^2 u^\delta}{\partial\varphi^\delta\partial\zeta^\delta}
		\right),\label{eq:sec03:computational2}\\
		%%%%%%%%%%
		&\frac{\partial v^\delta}{\partial\tau^\delta}+\zeta^\delta_t\frac{\partial v^\delta}{\partial\zeta^\delta}+\frac{\partial (v^\delta u^\delta)}{\partial\xi^\delta}+\zeta^\delta_x\frac{\partial(v^\delta u^\delta)}{\partial\zeta^\delta}
		+\frac{\partial(v^\delta v^\delta)}{\partial \varphi^\delta}\notag\\
		&+\zeta^\delta_y\frac{\partial(v^\delta v^\delta)}{\partial\zeta^\delta}
		+\zeta^\delta_z\frac{\partial(v^\delta w^\delta)}{\partial\zeta^\delta}
		=-\frac{1}{\rho^\delta}\left(\frac{\partial p^\delta}{\partial \varphi^\delta}+\zeta^\delta_y\frac{\partial p^\delta}{\partial \zeta^\delta}\right)\notag\\
		&+\nu^\delta\left(
		\frac{\partial^2 v^\delta}{\partial\xi^{^\delta2}}+\frac{\partial^2 v^\delta}{\partial\varphi^{\delta2}}+\left(\zeta_x^{\delta2}+\zeta_y^{\delta2}+\zeta_z^{\delta2}\right)\frac{\partial^2 v^\delta}{\partial\zeta^{\delta2}}\right.\notag\\
		&\left.+2\zeta_x^\delta\frac{\partial^2 v^\delta}{\partial\xi^\delta\partial\zeta^\delta}+2\zeta_y\frac{\partial^2 v^\delta}{\partial\varphi^\delta\partial\zeta^\delta}
		\right),\label{eq:sec03:computational3}\\
		%%%%%%%%%%%
		&\frac{\partial w^\delta}{\partial\tau^\delta}+\zeta^\delta_t\frac{\partial w^\delta}{\partial\zeta^\delta}+\frac{\partial (w^\delta u^\delta)}{\partial\xi^\delta}+\zeta^\delta_x\frac{\partial(w^\delta u^\delta)}{\partial\zeta^\delta}
		+\frac{\partial(w^\delta v^\delta)}{\partial \varphi^\delta}\notag\\
		&+\zeta^\delta_y\frac{\partial(w^\delta v^\delta)}{\partial\zeta^\delta}
		+\zeta^\delta_z\frac{\partial(w^\delta w^\delta)}{\partial\zeta^\delta}
		=-\frac{1}{\rho^\delta}\zeta^\delta_z\frac{\partial p^\delta}{\partial \zeta^\delta}\notag\\
		&+\nu^\delta\left(
		\frac{\partial^2 w^\delta}{\partial\xi^{^\delta2}}+\frac{\partial^2 w^\delta}{\partial\varphi^{\delta2}}+\left(\zeta_x^{\delta2}+\zeta_y^{\delta2}+\zeta_z^{\delta2}\right)\frac{\partial^2 w^\delta}{\partial\zeta^{\delta2}}\right.\notag\\
		&\left.+2\zeta_x^\delta\frac{\partial^2 w^\delta}{\partial\xi^\delta\partial\zeta^\delta}+2\zeta_y\frac{\partial^2 w^\delta}{\partial\varphi^\delta\partial\zeta^\delta}
		\right),\label{eq:sec03:computational4}
	\end{align}
\end{subequations}
\textcolor{black}{where $(u,v,w)$ denotes the velocity vector in the computational domain, and the superscript $\delta=(a,w)$ indicates variables in the air domain and the water domain, respectively.}
As detailed in \cite{li2022principal}, the governing equations are synchronously solved in the computational air and water domains using a fractional step method.
Through the boundary conditions on the air--water interface at each timestep, the air domain simulation transfers the pressure and shear stress at the water surface to the water domain, while the water domain simulation transfers the wave geometry and velocity to the air domain.
The matching of velocities and stresses between the air and water domains is enforced via an efficient iteration scheme.  
Our numerical solver has been extensively validated and used in many previous studies \citep[e.g.,][]{yang2010direct,yang2013dynamic,hao2019wind,xuan2019conservative,cao2020simulation,cao2021numerical,xuan2022analyses}.
\textcolor{black}{We note that the boundary-fitted numerical simulation method is more capable of resolving small scale surface deformations in the wind-wave generation process, compared with other two-phase flow simulation methods such as the volume-of-fluid (VOF) method.  In the boundary-fitted method, the dynamic and kinematic boundary conditions on the air--water interface are explicitly calculated.  However, in the VOF method, solving the transport equation for fraction function and reconstructing of the air--water interface are both diffusive.}

We conduct a DNS of the coupled air--water system in which the Reynolds number is limited owing to the computational cost of a wave-fitted grid simulation.
The air friction velocity is chosen as $u_\tau^a=\sqrt{\tau_s/\rho^a}=0.08\,\mathrm{m/s}$, with $\tau_s$ denoting the mean shear stress exerted on the air--water interface when the air--water interface is flat.  
\textcolor{black}{As shown in \citet{lin2008direct}, the shear stress on the air--water interface is in a quasi-steady state during the early stage of wind-wave generation process and will increase when the dominant waves are onset in the late stage.}
The height of both the air domain and the water domain is $H={0.0489}\,\mathrm{m}$.
\textcolor{black}{The streamwise domain length is $0.307\,\mathrm{m}$.}
The friction Reynolds number is $\mathit{Re}_\tau^a={u_\tau^aH}/{\nu^a}=268$ in the air domain and $\mathit{Re}_\tau^w={u_\tau^wH}/{\nu^w}=120$ in the water domain.
We choose realistic values both for the density and viscosity of air and water and for the surface tension of the air--water interface.
The capillary length scale is $l_c=\sqrt{\sigma/(\rho^wg)}=0.056H$ in the present study, and the surface tension of the air--water interface is $\sigma=7.35\times 10^{-2}\, \mathrm{N\,m^{-1}}$.
A grid of $128^3$ is adopted to conduct an ensemble simulation of the wind-wave generation problem.
The total number of simulations is $10$.
The key simulation parameters are summarized in \cref{tab:parameter}.  
\textcolor{black}{We note that in the present study, we perform ensemble simulations of wind-wave generation process under a moderate wind condition.  The Reynolds number scaling of pressure and shear stress fluctuations in wall-bounded turbulence suggests that the physical mechanisms of wind-wave generation discussed in this paper are also valid in a variety of wind conditions.  However, a quantitative study of the dependency of wave growth rate on the wind speed should be a topic of future research.}

\begin{table}[H]
	\caption{Summary of the simulation parameters. The superscript `$+$' denotes normalization by the viscous length scale $\nu^a/u_\tau^a$ in the air domain.}
	\begin{center}
		\begin{tabular}{|c|c|c|c|c|c|}
			\hline
			$Re_\tau^a$&$Re_\tau^w$& $\Delta x^+$ & $\Delta y^+$ & $\Delta z^+_{min}$ & $\Delta z^+_{min}$\\
			\hline
			268&120& 13.1 & 6.6 &0.234 & 3.7 \\
			\hline
		\end{tabular}
		\label{tab:parameter}
	\end{center}
\end{table}

\section{High-order surface elevation statistics}\label{sec:high-order_statistics}
The principal stage (the surface elevation variance grows linearly with time), the transition stage, and the exponential growth stage, which cover the entire wind-wave generation period, can be explicitly captured in the DNS \citep{li2022principal}. Based on the analysis of the surface elevation variance $\langle\eta^2\rangle$, \citep{li2022principal} classified the principal stage as occurring when $tu_\tau^a/H<20$, transition stage as occurring when $25<tu_\tau^a/H<45$, and the exponential stage as occurring when $tu_\tau^a/H>50$.  
In this section, we further analyze the evolution of the high-order surface elevation statistics and explore their characteristics in different wind-wave generation stages.

We first focus on the evolution of the moments of the absolute surface elevation value, defined as $\langle|\eta|^n\rangle$ for any positive integer $n$.  When $n$ equals $2$, $\langle|\eta|^n\rangle$ becomes the surface elevation variance, which was systematically studied in \citet{li2022principal}.
In the principal stage, the surface elevation variance grows linearly with time, i.e., $\langle\eta^2\rangle\sim t$.  This argument can be directly generalized to the case of arbitrary $n$:
\begin{align}
	\langle |\eta|^n \rangle \sim t^{\frac{n}{2}},\quad\quad \forall n\in \mathbb{Z}^+,\label{eq:moment_eta}
\end{align}
where ${\mathbb Z}^+$ denotes the space of positive integers.  We note that in the framework of the Phillips theory \citep{phillips1957generation}, the governing equation of the surface elevation $\eta$ is a second-order stochastic differential equation in Fourier space:
\begin{align}
	\frac{\partial^2 \hat{\eta}}{\partial t^2}+\Lambda^2(\bm k)\hat\eta=-\frac{k}{\rho^w}\hat{p}(\bm k ,t),\label{eq:eta}
\end{align}
where $\widehat{(\,\cdot\,)}$ denotes variables in Fourier space, $k$ denotes the modulus of the wavenumber $\bm k$, and $\Lambda(\bm k)$ denotes the dispersion relation of surface waves.
The solution of the surface elevation $\eta$ can be explicitly calculated in \cref{eq:eta}.
\cite{li2022principal} provided a rigorous but sophisticated derivation for a quantitative expression of \cref{eq:moment_eta} when $n=2$.
For other integers $n$, a quantitative expression of \cref{eq:moment_eta} can be obtained via similar techniques, which may be the focus of future research.
In the present study, we focus on the numerical evidence supporting the hypothesis that \cref{eq:moment_eta} holds in the principal stage.

\begin{figure}[H]
	\centering
	\includegraphics[trim=0.3in  0.3in 0.1in 0,clip,width=1.0\columnwidth]{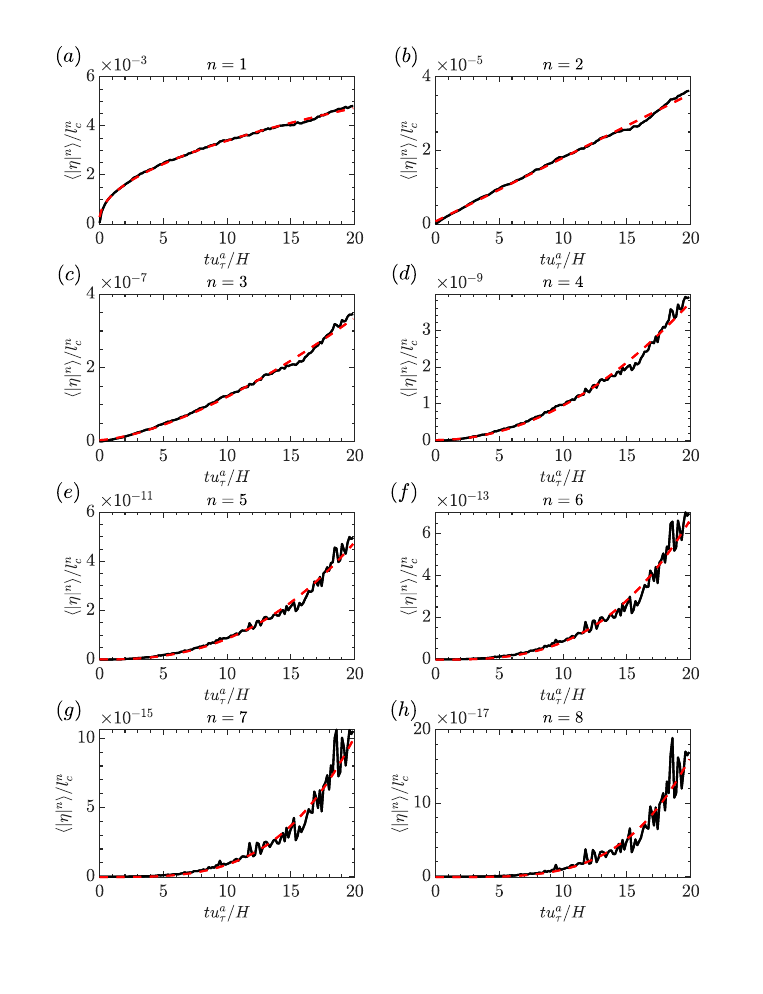}
	\caption{($a$)--($h$): Evolution of the wave statistics $\langle|\eta|^n\rangle$ for $n=1\dots8$ in the principal stage.  Black solid lines denote the DNS results, while red dashed lines denote the fitted curve of $A_1t^{\frac{n}{2}}+A_2$.} \label{fig:wave_pnorm}
\end{figure}

\Cref{fig:wave_pnorm} shows the evolution of the surface elevation moment $\langle|\eta|^n\rangle$ for $n=1\dots8$ normalized by the capillary length scale $l_c$ in the principal stage (when $tu_\tau^a/H<20$).  The black lines in this figure denote the DNS results, while the red dashed lines denote the fitted curve of a power-law function, $A_1t^{n/2}+A_2$, where $A_1$ and $A_2$ are constants to be determined.  The coefficient of determination (i.e., $R^2$) exceeds 0.95 for all cases, indicating that the DNS results of the surface elevation moment match well with \cref{eq:moment_eta}.

\textcolor{black}{During the early stage of wind-wave generation process, the wave nonlinearity is negligible, and the surface elevation statistics can be approximated as a Gaussian distribution at each time instant.  Under the Gaussian distribution hypothesis, the high-order moments $\langle |\eta|^n\rangle$ can be derived as 
	\begin{align}
		\langle |\eta|^n\rangle = s^n\frac{2^{\frac{n}{2}}\Gamma(\frac{n+1}{2})}{\sqrt{\pi}},\label{eq:gassian}
	\end{align}
	where $s$ denotes the standard deviation and $\Gamma$ denotes the gamma function.  Equation~\eqref{eq:gassian} indicates a power-law scaling of the standard deviation $s$ on the high-order moments $\langle |\eta|^n\rangle $.  Therefore, the linear growth of the second-order moment $\langle\eta^2\rangle$ in the Phillips theory and the Gaussian distribution hypothesis of the surface elevation $\eta$ can result in the temporal growth behavior of high-order moments $\langle |\eta|^n\rangle$ in Equation~\eqref{eq:moment_eta}.  Nevertheless, a rigorous justification of the Gaussian distribution hypothesis in the Phillips theory is an open question. 
}

The validation of \cref{eq:moment_eta} can benefit theoretical studies on other wave statistics of interest, such as the surface elevation skewness and kurtosis, which measure the asymmetry and randomness of surface elevations, respectively. For an instantaneous wavefield, the surface elevation skewness $C_3$ and kurtosis $C_4$ are defined as
\begin{align}
	C_3&=\frac{\langle\eta^3\rangle}{\langle\eta^2\rangle^{3/2}},\\
	C_4&=\frac{\langle\eta^4\rangle}{\langle\eta^2\rangle^{2}}.
\end{align}
\textcolor{black}{Here, the operator $\langle\,\cdot\,\rangle$ denotes the spatial and ensemble average. }

From the scaling behavior in \cref{eq:moment_eta}, we have the following estimation in the principal stage:
\begin{align}
	&|C_3|<\frac{\langle|\eta|^3\rangle}{\langle\eta^2\rangle^{3/2}}\sim \frac{t^{3/2}}{t^{3/2}}= 1,\label{eq:est_C3}\\
	&C_4\sim \frac{t^2}{t^2}= 1.\label{eq:est_C4}
\end{align}
\Cref{eq:est_C3} indicates that the magnitude of the skewness $C_3$ does not grow with time in the principal stage of wind-wave generation; likewise, \cref{eq:est_C4} indicates that the kurtosis is constant in the principal stage.

\begin{figure}[H]
	\centering
	\includegraphics[trim=0.0in  0.0in 0.0in 0,clip,width=0.97\columnwidth]{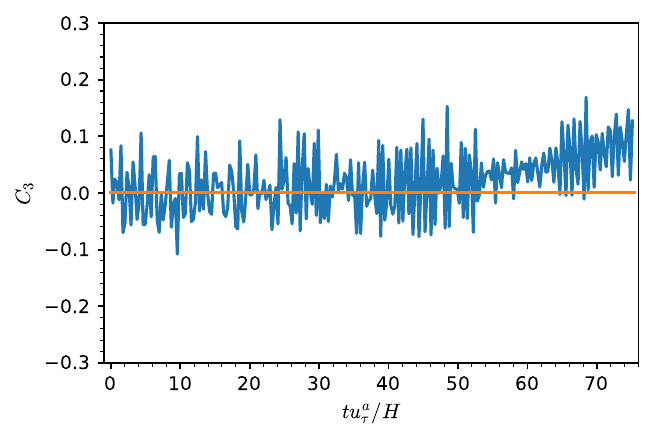}
	\caption{Evolution of the surface elevation skewness $C_3$.} \label{fig:wave_c3}
\end{figure}

\begin{figure}[H]
	\centering
	\includegraphics[trim=0.0in  0.0in 0.0in 0,clip,width=0.97\columnwidth]{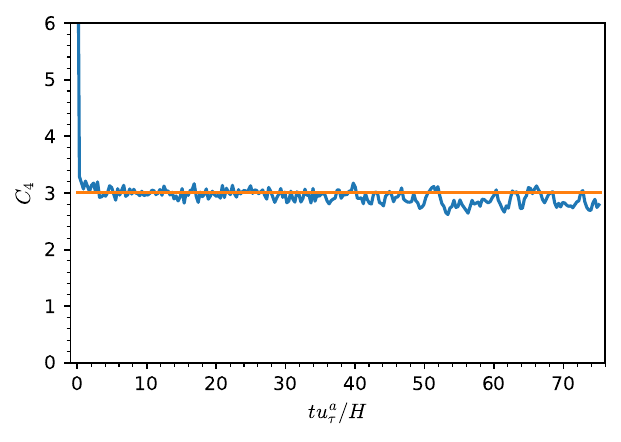}
	\caption{Evolution of the surface elevation kurtosis $C_4$.} \label{fig:wave_c4}
\end{figure}

Furthermore, \Cref{fig:wave_c3} and \Cref{fig:wave_c4} show the DNS results for the evolution of the surface elevation skewness $C_3$ and kurtosis $C_4$ with time, respectively.
As expected, in the principal stage (when $tu_\tau^a/H<20$), the skewness $C_3$ oscillates near 0, and the kurtosis $C_4$ oscillates near 3.
Neither $C_3$ nor $C_4$ grows with time in the principal stage, which agrees with the above analysis.
The values of $C_3=0$ and $C_4=3$ represent a Gaussian distribution, which could reasonably describe the randomness of surface elevations in the principal stage of wind-wave generation.
In the late stage (e.g., $tu_\tau^a/H>50$), when the dominant waves are onset and the surface elevation variance grows exponentially with time, the skewness exceeds zero and grows with time, and the kurtosis is slightly below 3.
Positive values of the skewness $C_3$ indicate that the surface waves are nonlinear, and the surface elevation profiles shift from a sinusoidal shape to a Stokes wave.
Values of the kurtosis $C_4<3$ indicate that a limited number of dominant waves are generated in the exponential stage and that the effect of nonlinear wave--wave interactions is weak.
For instance, the kurtosis of a single sinusoidal wave is $C_4=1.5$, and the kurtosis satisfies $C_4>3$ in a broadband wavefield where nonlinear wave--wave interactions dominate the transfer of energy \citep[see][]{longuet1963effect}.
These behaviors of the skewness $C_3$ and kurtosis $C_4$ are in agreement with experimental studies of young wavefields \cite[e.g.,][]{hatori1984nonlinear,zavadsky2013statistical,zavadsky2017water}.

\begin{figure}[H]
	\centering
	\includegraphics[width=1.0\columnwidth]{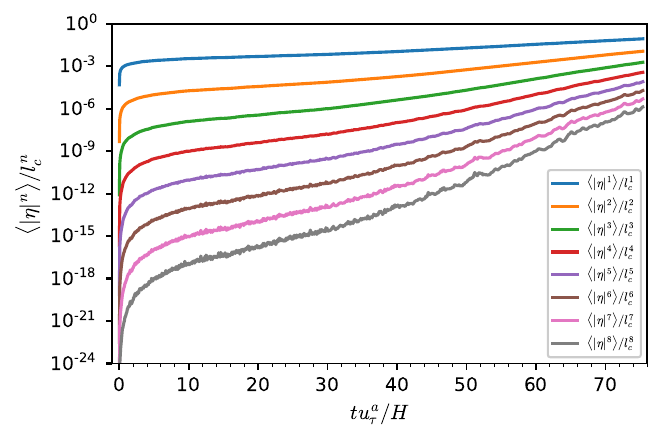}
	\caption{Evolution of the wave statistics $\langle|\eta|^n\rangle$ in the entire wind-wave generation process.  The $y$-axis is plotted on a logarithmic scale.} \label{fig:wave_pnorm2}
\end{figure}

Next, we investigate the behavior of the surface elevation moment $\langle|\eta|^n\rangle$ in the exponential growth stage.  \Cref{fig:wave_pnorm2} shows the DNS results of $\langle|\eta|^n\rangle$ for $n=1\dots8$ normalized by the capillary length scale $l_c$ in the entire wind-wave generation process.
Note that the $y$-axis in \cref{fig:wave_pnorm2} is plotted on a logarithmic scale; thus, a straight line in this plot indicates exponential growth with time.
As shown in \cref{fig:wave_pnorm2}, the surface elevation moments $\langle|\eta|^n\rangle$ for $n=1\dots8$ grow exponentially with time, but their exponential growth rates differ.

\begin{figure}[H]
	\centering
	\includegraphics[width=1.0\columnwidth]{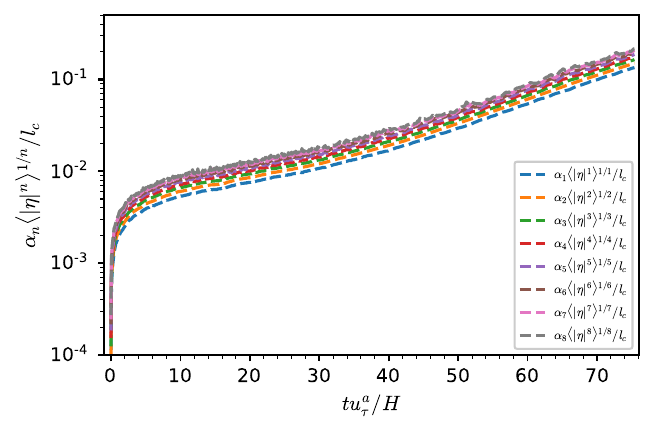}
	\caption{Evolution of the wave statistics $\langle|\eta|^n\rangle^{1/n}$ normalized by the factor $\alpha_n$ in the entire wind-wave generation process.} \label{fig:wave_pnorm_logscale_profile}
\end{figure}

In the exponential growth stage, we can model the surface elevation evolution as the summation of multiple single wave trains, the amplitudes of which grow exponentially with time.
For simplicity, we adopt the following estimation:
\begin{align}
	\eta=\exp\left(\frac{1}{2}\beta \omega t\right)\cos(kx-\omega t),\label{eq:exp_model_eta}
\end{align}
where the growth rate parameter $\beta$ is defined as in \cref{eq:beta}.
Under this estimation, we can obtain a universal profile of high-order moments: $\alpha_n\langle{|\eta|^n}\rangle^{1/n}/l_c$.
By direct calculation, the constant coefficient $\alpha_n$ is
\begin{align}
	\alpha_n=\left(\frac{\sqrt{\pi}\Gamma\left(\frac{n+2}{2}\right)}{\Gamma\left(\frac{n+1}{2}\right)}\right)^{1/n},
\end{align}
where $\Gamma(z)$ is the gamma function defined as $\Gamma(z)=\int_0^\infty x^{z-1}e^{-x}\mathrm{d}x$.
The profiles of $\alpha_n\langle{|\eta|^n}\rangle^{1/n}/l_c$ collapse for all $n$ in the exponential growth stage.
Moreover, \Cref{fig:wave_pnorm_logscale_profile} shows the DNS results of $\alpha_n\langle{|\eta|^n}\rangle^{1/n}/l_c$ for $n=1\dots8$, and all the profiles exhibit a universal shape.
We note that the slight deviations of the positions among the different DNS profiles are due to the assumption of only one dominant wave defined in \cref{eq:exp_model_eta}.
In the DNS of \citet{li2022principal}, the top three wave components have similar wave energy in the exponential stage.
Thus, introducing more wavenumbers into \cref{eq:exp_model_eta} could improve the accuracy of modeling the universal profile of high-order moments in the exponential growth stage.

\section{Anisotropy factor of the surface elevation}\label{sec:shape_factor}
\begin{figure}[H]
	\centering
	\includegraphics[trim=0.0in  0.0in 0.0in 0,clip,width=1.0\columnwidth]{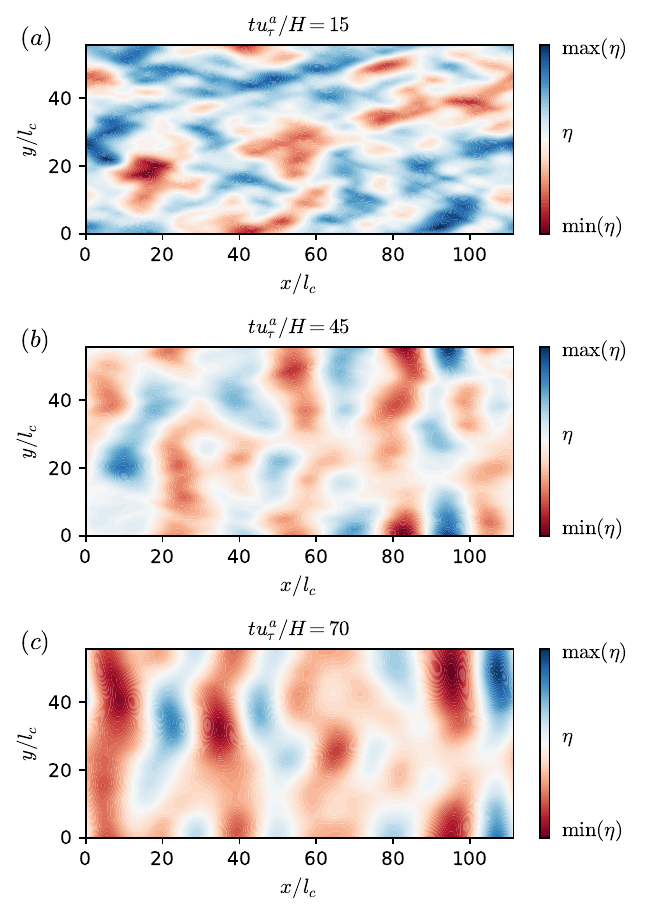}
	\caption{Contours of the instantaneous surface elevation at $(a)$: $tu_\tau^a/H=15$, $(b)$: $tu_\tau^a/H=45$, and $(c)$: $tu_\tau^a/H=70$.} \label{fig:wave_field}
\end{figure}

In the previous section, we investigate the evolution of the high-order surface elevation moments in the wind-wave generation process.
However, spatial inhomogeneity is not taken into account in the analysis of these high-order surface elevation moments.
In previous DNS studies \citep{lin2008direct,li2022principal}, the surface elevations exhibited distinct spatial characteristics in different wind-wave generation stages.
\Cref{fig:wave_field} visualizes the instantaneous surface elevations at different time instants.
As shown in \cref{fig:wave_field}(\textit{a}), the surface elevations exhibit streak-like structures, and the variations in the surface elevation along the $y$-axis are obvious.
Subsequently, as dominant waves are generated (see \cref{fig:wave_field}\textit{b} and \textit{c}), the variations in the surface elevation along the $y$-axis decrease with time, and regular waves propagating in the wind direction can be observed.

We introduce an anisotropy factor of the surface elevation to quantify the spatial inhomogeneity in the entire wind-wave generation process.
For an instantaneous wavefield $\eta(x,y)$, the anisotropy factor $F$ is defined as 
\begin{align}
	F=\frac{\langle\sigma_y(\eta)\rangle_x}{\sigma_{xy}(\eta)},\label{eq:def_F}
\end{align}
where $\langle\,\cdot\,\rangle_x$ denotes a spatial averaging operation in the $x$-direction, $\sigma(\,\cdot\,)_y$ denotes the standard deviation in the $y$-direction, and $\sigma(\,\cdot\,)_{xy}$ denotes the standard deviation in both the $x$- and the $y$-directions.

It is straightforward to show that the anisotropy factor $F$ varies within the interval $[0,1]$.
$F$ equals 0 if $\eta$ is homogeneous in the $y$-direction, and $F$ is close to $1$ when the variation in $\eta$ in the $y$-direction is significantly larger than that in the $x$-direction.
Therefore, the anisotropy factor $F$ can quantify the spatial inhomogeneity of the wind-generated wavefield.

\begin{figure}[H]
	\centering
	\includegraphics[trim=0.0in  0.0in 0.0in 0,clip,width=1.0\columnwidth]{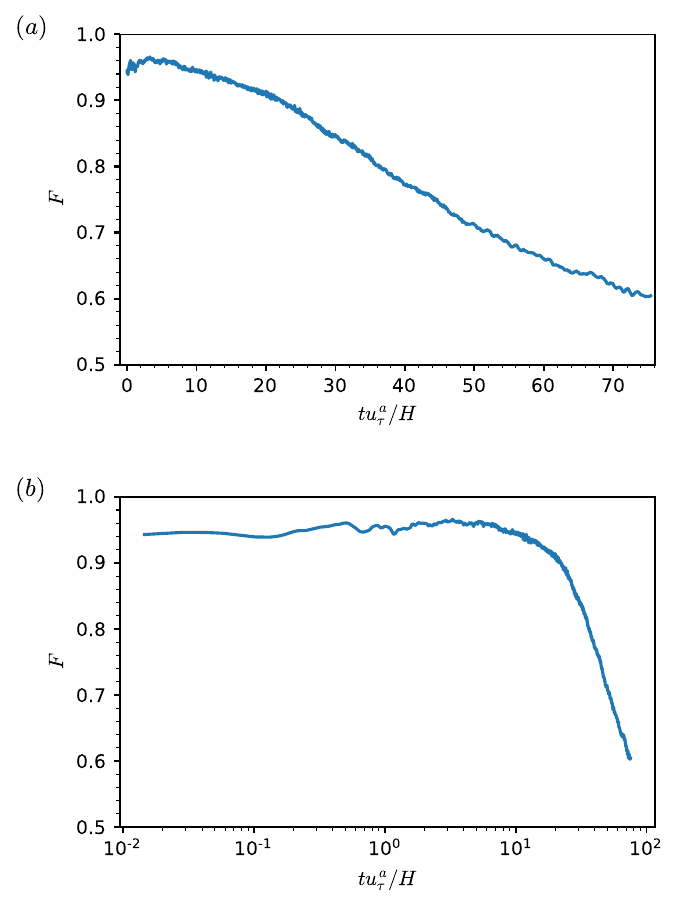}
	\caption{Evolution of the surface elevation anisotropy factor $F$ defined in \cref{eq:def_F}.  Panel $(a)$ plots both the $x$- and the $y$-axes on a normal scale, while panel $(b)$ plots the $x$-axis on a logarithmic scale and the $y$-axis on a normal scale.} \label{fig:wave_shape_factor}
\end{figure}

\Cref{fig:wave_shape_factor} shows the DNS results for the evolution of the anisotropy factor $F$ defined in \cref{eq:def_F}.
As shown in \cref{fig:wave_shape_factor}(\textit{a}), $F$ is close to $1$ in the early stage of wind-wave generation.
$F$ decreases with time into the late stage, indicating that the surface elevations along the $y$-direction grow smoothly.
The evolution of the anisotropy factor $F$ is consistent with the observed instantaneous surface elevations.

In addition, the anisotropy factor $F$ also quantifies the shape-preserving features of the surface elevation.
In the Phillips theory \citep{phillips1957generation}, the shape of the wave energy spectrum does not vary with time in the principal stage;
therefore, the anisotropy factor $F$ remains constant in the principal stage.
We examine this corollary of the Phillips theory in \cref{fig:wave_shape_factor}(\textit{b}), where the $x$-axis is plotted on a logarithmic scale.
The anisotropy factor $F$ reaches a plateau during the time interval $tu_\tau^a/H\in[0,10]$, indicating that this shape-preserving feature exists in the principal stage.
We note that the duration of the principal stage is defined as $tu_\tau^a/H\in[0,20]$ based on the analysis of the surface elevation variance $\langle\eta^2\rangle$ in \citet{li2022principal}.
In the present study, we show that this shape-preserving feature is violated when $tu_\tau^a/H>10$ even though the linear growth behavior of the surface elevation variance is still reasonably preserved.

\section{Correlation function}\label{sec:correlation_function}
In this section, we investigate the correlation function between the surface elevation and surface turbulent pressure fluctuations.
Energy is transferred between the airflow and wavefield through the work done by the air pressure on the air--water interface.
Accordingly, understanding the correlations between surface elevations and pressure fluctuations can provide new insights into the energy transfer behaviors in different wind-wave generation stages.
In particular, we focus on the temporal evolution of the correlation function between the surface elevation gradient $\eta_x$ and pressure fluctuation $p$, i.e., $\mathrm{Corr}(\eta_x,p)$, which is closely related to the form drag.  The correlation function $\mathrm{Corr}(\eta_x,p)$ is defined as
\begin{align}
	\mathrm{Corr}(\eta_x,p)=\frac{\langle\eta_xp\rangle}{\langle\eta_x^2\rangle^{1/2}\langle p^2\rangle^{1/2}}.
\end{align}
Next, we investigate the behavior of $\mathrm{Corr}(\eta_x,p)$ with a theoretical approach.  The evolution of the surface elevation $\eta$ in the principal stage can be obtained by the direct calculation of \cref{eq:eta}, and the expression of $\eta$ in Fourier space is 
\begin{align}
	\hat{\eta}=\frac{k}{\rho^w\Lambda}\int_0^t\sin(\Lambda\tau-\Lambda t)\hat{p}(\bm k,t)\mathrm{d}\tau.
\end{align}
The term $\langle\eta_xp\rangle$ can be rewritten in the Fourier space owing to the Plancherel theorem:
\begin{align}
	\langle\eta_xp\rangle=L\langle\left(\widehat{\eta_x}\right)^*\hat{p}\rangle,
\end{align}
where $(\,\cdot\,)^*$ denotes the complex conjugate and $L$ is a universal constant related to the definition of the Fourier transform.  By calculating $\left(\widehat{\eta_x}\right)^*\hat{p}$, we can obtain
\begin{align}
	\left(\widehat{\eta_x}\right)^*\hat{p}=\frac{-\mathrm{i}k_xk}{\rho^w\Lambda}\int_0^t\sin(\Lambda\tau-\Lambda_t)\hat{p}^*(\bm k,\tau)\hat{p}(\bm k,t)\mathrm{d}\tau. \label{eq:eta_x_p}
\end{align}
Taking the expected value on both sides of \cref{eq:eta_x_p}, we obtain 
\begin{align}
	\mathrm{E}\left[\left(\widehat{\eta_x}\right)^*\hat{p}\right]=\frac{-\mathrm{i}k_xk}{\rho^w\Lambda}\int_0^t\sin(\Lambda\tau-\Lambda t)\widehat\Pi_p(\bm k, t-\tau)\mathrm{d}\tau,
\end{align}
where $\widehat\Pi_p(\bm k, \tau)$ denotes the spatial Fourier transform of the two-point, two-time autocovariance of pressure fluctuations $\Pi_p(\bm x, \bm t)$.
$\widehat\Pi_p(\bm k, \tau)$ is positive for any $\tau$ and $\bm k$ and satisfies the following property:
\begin{align}
	\int_{-\infty}^{\infty}\widehat\Pi_p(\bm k, \tau)\mathrm{d}\tau<\infty.
\end{align}
Therefore, we have the following estimation:
\begin{align}
	\left|\mathrm{E}\left[\left(\widehat{\eta_x}\right)^*\hat{p}\right]\right|&
	<\frac{\left|k_xk\right|}{\rho^w\Lambda}\int_0^t|\sin(\Lambda\tau-\Lambda t)|\widehat\Pi_p(\bm k, t-\tau)\mathrm{d}\tau\notag\\
	&<\frac{\left|k_xk\right|}{\rho^w\Lambda}\int_0^t\widehat\Pi_p(\bm k, t-\tau)\mathrm{d}\tau\notag\\
	&<\frac{\left|k_xk\right|}{\rho^w\Lambda}\int_{-\infty}^\infty\widehat\Pi_p(\bm k, \tau)\mathrm{d}\tau.\label{eq:est_corr}
\end{align}
\Cref{eq:est_corr} indicates that the magnitude of $\mathrm{E}\left[\left(\widehat{\eta_x}\right)^*\hat{p}\right]$ is bounded by a function related only to the wavenumber $\bm k$.
Therefore, we conclude that the term $\langle\eta_xp\rangle$ does not depend on the time $t$.
Moreover, because the surface elevation slope variance $\langle\eta_x^2\rangle$ is linearly dependent on $t$ but the pressure fluctuation variance $\langle\eta_xp\rangle$ is not a function of $t$, we obtain the following estimation:
\begin{align}
	|\mathrm{Corr}(\eta_x,p)|\sim t^{-\frac{1}{2}}.\label{eq:corr_t}
\end{align}
\Cref{eq:corr_t} shows that the correlation function $\mathrm{Corr}(\eta_x,p)$ decays with time as $t^{-1/2}$ in the principal stage.

\begin{figure}[H]
	\centering
	\includegraphics[trim=0.0in  0.0in 0.0in 0,clip,width=1.0\columnwidth]{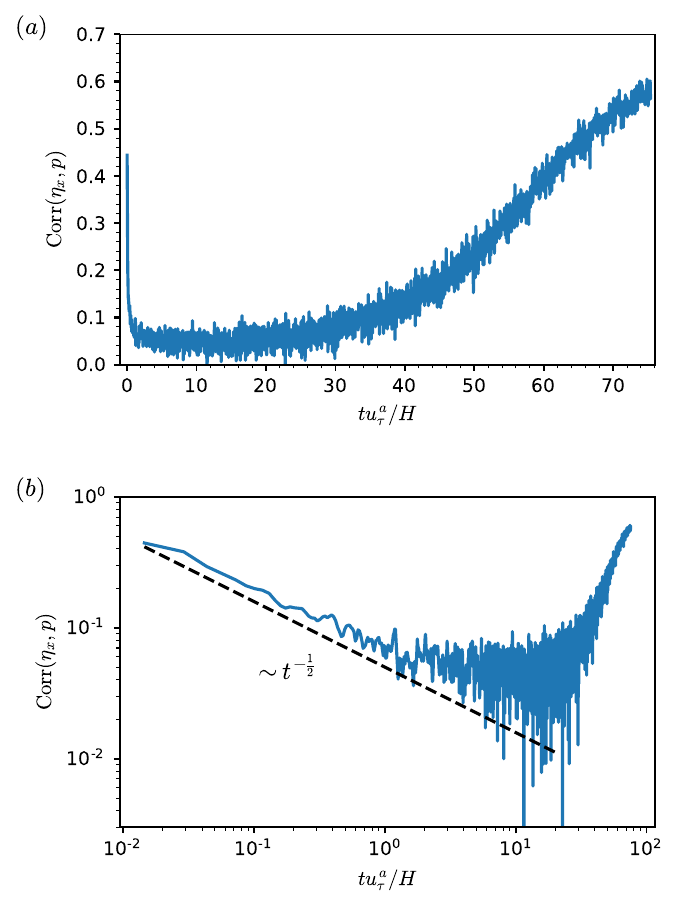}
	\caption{Evolution of the correlation function $\mathrm{Corr}(\eta_x,p)$.  Panel $(a)$ plots both the $x$- and the $y$-axes on a normal scale, while panel $(b)$ plots both the $x$- and the $y$-axes on a logarithmic scale.} \label{fig:corr_etax_p}
\end{figure}

Next, we evaluate the above theoretical analysis using the DNS results.  \Cref{fig:corr_etax_p} shows the evolution of the correlation function $\mathrm{Corr}(\eta_x,p)$ during the wind-wave generation process.  As shown in \cref{fig:corr_etax_p}(\textit{a}), $\mathrm{Corr}(\eta_x,p)$ first decays in the early stage and increases in the late stage. Positive values of $\mathrm{Corr}(\eta_x,p)$ indicate a positive drag force, which transfers energy from the airflow to the wavefield. \Cref{fig:corr_etax_p}(\textit{b}) illustrates the evolution of the correlation function $\mathrm{Corr}(\eta_x,p)$ in a log--log plot.  A power-law behavior with the index of $-1/2$ can be observed, which matches well with the result of our theoretical analysis via \cref{eq:corr_t}.
\textcolor{black}{The increase of the correlation function $\mathrm{Corr}(\eta_x,p)$ in the late stage indicates the occurrence of wave-induced air pressure fluctuation when the streamwise-traveling dominant waves have been generated.  The synchronization between the wave field and air pressure fluctuation triggers the critical layer instability and leads to an exponential growth of wave height \citep{miles1957generation}.}

\section{Conclusions}\label{sec:conclusions}
In the present study, we analyzed the wave characteristics for the classic wind-wave generation problem based on the DNS database of \cite{li2022principal}.
Using a combined numerical and theoretical approach, we studied the evolution of the high-order moments of the surface elevation magnitude $\langle|\eta|^n\rangle$, the anisotropy factor of the surface elevation $F$, and the correlation function between surface elevation gradients and pressure fluctuations $\mathrm{Corr}(\eta_x,p)$.
The linear growth behavior of the surface elevation variance was generalized to a power-law growth behavior with time for the high-order moments of the surface elevation magnitude $\langle|\eta|^n\rangle$ in the principal stage of wave development, and this outcome was validated using the DNS results. A universal profile of $\langle|\eta|^n\rangle$ was found in the exponential growth stage of wave development.  The skewness $C_3$ and kurtosis $C_4$ of surface elevations were analyzed. A theoretical analysis indicated that $C_3$ and $C_4$ remain constant in the principal stage of wave development, and the evolution of both $C_3$ and $C_4$ in the late stage was further discussed.  We introduced the anisotropy factor $F$ to quantify the spatial inhomogeneity of the surface elevation and to analyze the shape-preserving feature in the principal stage of wave development.  Finally, the correlation function between surface elevation gradients and pressure fluctuations $\mathrm{Corr}(\eta_x,p)$ was studied in the framework of the Phillips theory, and the power-law decay of $\mathrm{Corr}(\eta_x,p)$ with time was derived theoretically, with the result demonstrating good agreement with the numerical results.

\section{Acknowledgments}
\label{SECacknowledgements}

\textcolor{black}{This work is supported by the Office of Naval Research (N00014-19-1-2139, Program Manager Dr. Peter Chang; N00014-20-1-2747, Program Manager Dr. Scott Harper).} The computational resources provided by the DoD High Performance Computing Modernization Program (HPCMP) are gratefully acknowledged.

\bibliographystyle{SNHstyle}
\bibliography{120_SNH_Li_Shen_Arxiv}
\end{multicols*}
\end{document}